\documentclass{jps-cp}
\usepackage{txfonts} 

\title{Status of the Theoretical Calculation of Nuclear Electric Dipole Moment}

\author{Nodoka \textsc{Yamanaka}$^{1,2}$
}

\inst{$^{1}$IPNO, CNRS-IN2P3, Univ. Paris-Sud, Universit\'e Paris-Saclay, 91406 Orsay Cedex, France\\
$^{2}$iTHES Research Group, RIKEN, Wako, Saitama 351-0198, Japan}

\email{yamanaka@ipno.in2p3.fr}

\abst{
The electric dipole moment is a very sensitive probe of CP violation beyond the standard model.
Light nuclei are particularly interesting since the CP violation may be enhanced by nuclear many-body effects.
In this proceeding, we present the current status of the theoretical evaluations of the electric dipole moment of light nuclei in the ab initio approach and in the cluster model. 
We add the preliminary result of the evaluation of the electric dipole moment of $^7$Li which is treated in a cluster model with a triton.
}

\kword{CP violation, electric dipole moment, light nuclei, cluster model
}

\begin{document}
\maketitle

\def\vc#1{\mbox{\boldmath $#1$}}

\section{Introduction}

The excess of matter over antimatter, or the baryon number asymmetry, was generated in the early Universe by a theory fulfilling Sakharov's criteria \cite{sakharov}.
One of the required conditions, the CP violation, is known to have a critical deficit in the standard model (SM) of particle physics.
The search for new sources of CP violation beyond the SM is currently one of the most important fundamental problems of particle physics to be solved.

One of the most sensitive experimental observables to the CP violation beyond the SM is the electric dipole moment (EDM) \cite{engeledmreview,yamanakabook,atomicedmreview,chuppreview}.
Recently, the measurement of the nuclear EDM is under discussion \cite{anastassopoulos,jedi}.
The nuclear EDM has many advantages.
Like the EDM of other systems, the SM contribution is known to be very small \cite{yamanakasmedm}.
It can also be measured with high accuracy using storage rings \cite{storage1,storage2}, being free from the suppression of the EDM due the screening by atomic electrons \cite{schiff}.

Theoretically, the nuclear EDM is also very attractive, since the effect of CP violation may be enhanced by the many-body effect \cite{devriesedmreview,yamanakanuclearedmreview}.
The nuclear structure of light nuclei can be handled with good accuracy either ab initio or in the cluster model \cite{clusterreview3}.
The importance of light nuclei is also emphasized by the suppression of the EDM of heavy nuclei due to the configuration mixing \cite{yoshinaga1,yoshinaga2}.

The nuclear EDM has two leading contributions, namely, the intrinsic nucleon EDM contribution, and the polarization of the nucleus by the CP-odd nuclear force.
The former one is not enhanced at the nuclear level due to the pairing and the nonrelativistic nature of nuclei, so we do not treat it in this proceeding.
Here we review the current status of the theoretical calculations of the EDM of light nuclei due to the polarization generated by the one-pion exchange CP-odd nuclear force.
We also present the preliminary result on the calculations of the EDM of $^7$Li.

In the next section, we present the nuclear interactions used in our discussion.
In Section \ref{sec:polarization}, we review the current results of the nuclear EDM with the relevant physics.
The last section is devoted to the summary.

\section{Model setup and interactions}

\subsection{The Bare $N-N$ interaction}

\begin{figure}[tbh]
\begin{center}
\includegraphics[width=8cm]{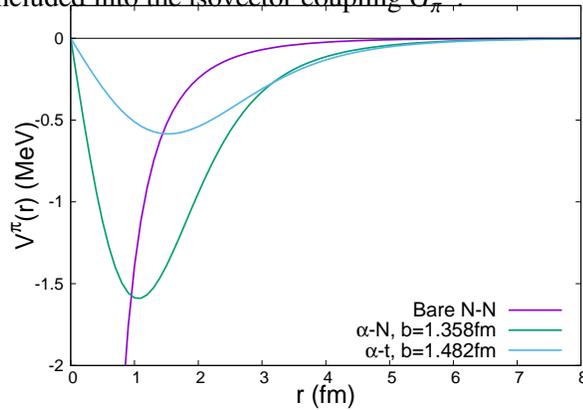}
\caption{
The radial shape of the one-pion exchange CP-odd nuclear force $V^\pi (r)$ for bare $N-N$ system, and in cluster models.
The coupling constant $\bar G_\pi^{(i)}\, (i=0,1,$ or $2)$ was factored out.
}
\label{fig:folding}
\end{center}
\end{figure}

The deuteron, $^3$He, and $^3$H are treated ab initio using the A$v$18 potential \cite{av18}.
The CP-odd nuclear force required to polarize the system is modeled by the one-pion exchange potential \cite{pvcpvhamiltonian3}.
\begin{eqnarray}
H_{P\hspace{-.35em}/\, T\hspace{-.5em}/\, }^\pi
& = &
\bigg\{ 
\bar{G}_{\pi}^{(0)}\,{\vc{\tau}}_{1}\cdot {\vc{\tau}}_{2}\, {\vc{\sigma}}_{-}
+\frac{1}{2} \bar{G}_{\pi}^{(1)}\,
( \tau_{+}^{z}\, {\vc{\sigma}}_{-} +\tau_{-}^{z}\,{\vc{\sigma}}_{+} )
+\bar{G}_{\pi}^{(2)}\, (3\tau_{1}^{z}\tau_{2}^{z}- {\vc{\tau}}_{1}\cdot {\vc{\tau}}_{2})\,{\vc{\sigma}}_{-} 
\bigg\}
\cdot
\frac{ \vc{r}}{r} \,
V(r)
,
\label{eq:CPVhamiltonian}
\end{eqnarray}
with the relative coordinate between the two nucleons denoted by $\vc{r} \equiv \vc{r}_1 - \vc{r}_2$.
We also define the spin and isospin matrices by ${\vc{\sigma}}_{-} \equiv {\vc{\sigma}}_1 -{\vc{\sigma}}_2$, ${\vc{\sigma}}_{+} \equiv {\vc{\sigma}}_1 + {\vc{\sigma}}_2$, ${\vc{\tau}}_{-} \equiv {\vc{\tau}}_1 -{\vc{\tau}}_2$, and ${\vc{\tau}}_{+} \equiv {\vc{\tau}}_1 + {\vc{\tau}}_2$.
Here we consider three isospin structures with independent CP-odd couplings $\bar G_\pi^{(i)}$ $(i=0,1,2)$.
The radial dependence of the one-pion exchange CP-odd nuclear force is given by
\begin{equation}
V^\pi (r)
= 
-\frac{m_\pi}{8\pi m_N} \frac{e^{-m_\pi r }}{r} \left( 1+ \frac{1}{m_\pi r} \right)
\ .
\end{equation}
Its shape is displayed in Fig. \ref{fig:folding}.
In the leading order of chiral perturbation, there are also additional interactions such as the short-range CP-odd $N-N$ interaction \cite{cpveft}, which will not be discussed due to the large uncertainty of the nuclear wave function at short distance \cite{bsaisou}, or the three-pion interaction \cite{cpveft2}, which can be effectively included into the isovector coupling $\bar G_\pi^{(1)}$.

\subsection{The cluster model}

It is known that the cluster model reproduces well the structure of light nuclei \cite{clusterreview3}.
In this study, we consider the $\alpha$-cluster model to describe the $^6$Li, $^9$Be, and $^{13}$C nuclei, and the $\alpha - ^3$H model for $^7$Li.
Regarding the CP-even interactions, we use the Kanada-Kaneko potential for the $\alpha -N$ system of $^6$Li and $^9$Be, and $^{13}$C \cite{kanada}.
For the $\alpha - \alpha$ system, we use the modified Hasegawa-Nagata potential \cite{hasegawa} needed for $^6$Li and $^9$Be, and the Schmid-Wildermuth potential \cite{schmid} for that of $^{13}$C.
For $\alpha -^3$H we use the interaction of Nishioka {\it et al}. \cite{nishioka}.
We exclude forbidden states using the Orthogonality Condition Model \cite{saito}.

To model the CP-odd potential between clusters, we use the folding \cite{horiuchi} of the bare CP-odd $N-N$ interaction of Eq. (\ref{eq:CPVhamiltonian}).
The oscillator parameter is taken as $b = 1.358$ fm and 1.482 fm for the CP-odd $\alpha - N$ and $\alpha - ^3$H potentials, respectively.
The shape of the above CP-odd interactions is shown in Fig. \ref{fig:folding}.
Note that the isoscalar and isotensor CP-odd nuclear forces vanish after folding since the spin and isospin shells are closed for the $\alpha$-clusters.

\section{Contribution of the CP-odd nuclear force to the nuclear electric dipole moment\label{sec:polarization}}

The polarization contribution to the nuclear EDM is defined by
\begin{eqnarray}
d_{A}^{\rm (pol)} 
&=&
\sum_{i=1}^{A} \frac{e}{2} 
\langle \, \tilde \Phi_J (A) \, |\, (1+\tau_i^3 ) \, r_{iz} \, | \, \tilde \Phi_J (A) \, \rangle
,
\label{eq:polarizationedm}
\end{eqnarray}
where $|\, \tilde \Phi_{J} (A)\, \rangle$ is the wave function of the nucleus $A$ polarized along the axis of the measurement.
The nuclear EDM can then be parametrized as a linear combination of CP-odd nuclear couplings:
\begin{eqnarray}
d_{A}^{\rm (pol)} 
&=&
\bar G_\pi^{(0)}
a_\pi^{(0)}
+\bar G_\pi^{(1)}
a_\pi^{(1)}
+\bar G_\pi^{(2)}
a_\pi^{(2)}
.
\label{eq:polarizationedm}
\end{eqnarray}
Interesting nuclei are thus those which have large coefficients $a_\pi^{(i)}$ $(i=0,1,2)$.

We summarize in Table \ref{table:nuclearedm} the current results of the calculations of nuclear EDMs.

\begin{table}[tbh]
\caption{
The EDM coefficients of the pion exchange CP-odd nuclear force.
The symbol ``$-$'' means that the coefficient cancels.
Those of the neutron are also given for comparison \cite{crewther}.
Note that the coefficients of light nuclei do not include the effect from the intrinsic nucleon EDM.
}
\label{table:nuclearedm}
\begin{center}
\begin{tabular}{l|ccc|}
  &$a_\pi^{(0)}$ ($10^{-2} e$ fm) & $a_\pi^{(1)}$ ($10^{-2} e$ fm) & $a_\pi^{(2)}$ ($10^{-2} e$ fm) \\ 
\hline
$n$ \cite{crewther} & $1$ & $- $ & $-1$  \\
$^{2}$H \cite{yamanakanuclearedm} & $-$ & $1.45 $ & $-$  \\
$^{3}$He \cite{yamanakanuclearedm}& $0.59$ & 1.08 & 1.68  \\
$^{3}$H \cite{yamanakanuclearedm} & $-0.59$ & 1.08 & $-1.70$  \\
$^{6}$Li \cite{yamanakanuclearedm}& $-$ & 2.2 & $-$  \\
$^{7}$Li & $-0.6$ & 1.6 & $-1.7$  \\
$^{9}$Be \cite{yamanakanuclearedm}  & $-$ & $1.4$ & $-$  \\
$^{13}$C \cite{c13edm} & $-$ & $-0.20 $ & $-$  \\
\hline
\end{tabular}
\end{center}
\end{table}

The results for the deuteron, $^3$He, and $^3$H are consistent with previous works \cite{liu,bsaisou}.
The deuteron is only sensitive to the isovector CP-odd nuclear force due to the isospin selection rule.
The EDMs of $^3$He and $^3$H are sensitive to all three isospin structures because of the isospin asymmetry.

The EDM of $^6$Li is only sensitive to the isovector coupling.
It receives contributions from the EDM of the deuteron subsystem and from the CP-odd $\alpha -N$ interaction.
Both only depends on the isovector CP-odd nuclear force.
It is also larger than the deuteron one.
This fact shows that the two contributions interfere constructively.
For $^9$Be, the EDM is smaller than that of $^6$Li, since the CP-odd $\alpha -N$ interaction is the only source of polarization.
The EDM of $^7$Li receives contribution from the EDM of the $^3$H cluster and from the CP-odd $\alpha -^3$H interaction, which also interfere constructively.
The results of the EDM of $^6$Li, $^7$Li and $^9$Be suggest an additive counting rule involving the EDM of clusters and the CP-odd $\alpha -N$ interaction with a contribution of $(0.005 - 0.007) \bar G_\pi^{(1)} e$ fm.

The polarization contribution to the EDM of $^{13}$C is however not respecting this counting rule.
The $^{13}$C nucleus has an opposite parity state at 3.1 MeV above the ground state, so we would expect a relatively large EDM, since the EDM is a mixing between states with opposite parity.
This state is however known to have a $^{12}$C cluster with different structure than that of the ground state \cite{yamada}.
Due to this bad overlap, the EDM of $^{13}$C is actually suppressed by one order of magnitude \cite{c13edm}.

\section{Summary\label{sec:summary}}

In this work, we made an overview of the current status of the nuclear EDM.
The study of $^6$Li, $^7$Li and $^9$Be revealed us that the EDM seems to obey a rough counting rule depending on the EDM of the cluster and the CP-odd $\alpha -N$ polarization.
The EDM of $^{13}$C is however not obeying it, and it has a suppressed coefficient, which is due to the bad overlap between even- and odd-parity states.
Those results tell us that each light nucleus has its own mechanism to generate its EDM, so the EDM of experimentally measurable light nuclei has to be evaluated independently.

\end{document}